# Automated PII Extraction from Social Media for Raising Privacy Awareness:
# A Deep Transfer Learning Approach

Yizhi Liu
*Department of Decision, Operations and Information Technologies*
University of Maryland
College Park, Maryland
yizhiliu@umd.edu

Fang Yu Lin
*Management Information Systems*
University of Arizona
Tucson, Arizona
fylin@email.arizona.edu

Mohammadreza Ebrahimi
*School of Information Systems and Management*
University of South Florida
Tampa, Florida
ebrahimim@usf.edu

Weifeng Li
*Department of Management Information Systems*
University of Georgia
Athens, Georgia
weifeng.li@uga.edu

Hsinchun Chen
*Management Information Systems*
University of Arizona
Tucson, Arizona
hchen@eller.arizona.edu

*Abstract*— Internet users have been exposing an increasing amount of Personally Identifiable Information (PII) on social media. Such exposed PII can be exploited by cybercriminals and cause severe losses to the users. Informing users of their PII exposure in social media is crucial to raise their privacy awareness and encourage them to take protective measures. To this end, advanced techniques are needed to extract users' exposed PII in social media automatically, whereas most existing studies remain manual. While Information Extraction (IE) techniques can be used to extract the PII automatically, Deep Learning (DL)-based IE models alleviate the need for feature engineering and further improve the efficiency. However, DL-based IE models often require large-scale labeled data for training, but PII-labeled social media posts are difficult to obtain due to privacy concerns. Also, these models rely heavily on pre-trained word embeddings, while PII in social media often varies in forms and thus has no fixed representations in pre-trained word embeddings. In this study, we propose the Deep Transfer Learning for PII Extraction (DTL-PIIE) framework to address these two limitations. DTL-PIIE transfers knowledge learned from publicly available PII data to social media in order to address the problem of rare PII-labeled data. Moreover, our framework leverages Graph Convolutional Networks (GCNs) to incorporate syntactic patterns to guide PIIE without relying on pre-trained word embeddings. Evaluation against benchmark IE models indicates that our approach outperforms state-of-the-art DL-based IE models. An ablation analysis further confirms the efficacy of each component in our model. Our proposed framework can facilitate various applications, such as PII misuse prediction and privacy risk assessment, thereby protecting the privacy of internet users.

## I. Introduction

In recent years, internet users have been sharing an increasing amount of personal information on social media [1]. The shared information often contains various Personally Identifiable Information (PII) of the users, such as their name, geolocation, birthdate, contact, and SSN [2]. Such PII can be easily exploited and misused by adversaries and cyber criminals, causing severe financial losses and reputation damage.

To raise users' awareness of their exposed PII in social media and associated privacy risks, a prevalent approach is to inform users of their extent of PII exposure in social media, which has been shown to be effective in helping users take protective measures [3]. However, this entails extracting users' exposed PII in social media automatically, which is a non-trivial task. Most existing methods manually investigate social media posts to estimate the PII exposure using heuristic rules. Given the enormous amount of social media content, these manual methods lack efficiency, calling for automated alternatives. Therefore, Information Extraction (IE) is needed to facilitate the automation of PII extraction. However, the types of PII attributes are diverse and developing feature engineering for various types of PII attributes can be time-consuming and labor intensive. As a result, Deep Learning (DL) approaches have been adopted to automatically learn deep representations of various PII attributes in medical notes [4], news [5], and the dark web [6]. Despite the significant achievements of DL approaches in IE, existing DL-based IE research has two limitations in extracting PII from social media. First, DL models often require large-scale labeled data for training, but PII-labeled social media posts are rare due to privacy concerns [1]. Second, most deep learning models rely heavily on pre-trained word embeddings. Nonetheless, PII in social media posts often varies in forms (e.g., arbitrary formats for phone numbers) and cannot be captured by the fixed pattern matching of pre-trained word embeddings.

In this study, we aim to extract users' exposed PII in social media automatically. This can facilitate various applications to raise users' privacy awareness, such as the prediction of PII misuse and privacy risk assessment. To achieve this goal, we propose the Deep Transfer Learning for PII Extraction (DTL-PIIE) framework. Our framework utilizes manually labeled public data as the source domain and transfers the knowledge to social media in order to address the lack of rare PII-labeled data. In particular, our framework focuses on transferring syntactic patterns learned by Graph Convolutional Networks (GCNs) from PII-labeled to guide the extraction of PII in online social media, without relying on pre-trained word embeddings for PII attributes.

## II. Literature Review

Three relevant areas of research are reviewed. First, we examine how previous studies analyze users' PII exposure on social media. Second, we review Deep Learning (DL) for Information Extraction (IE) literature to understand prevailing state-of-the-art DL-based IE methods to guide PII extraction from social media posts. Third, we identify Deep Transfer Learning (DTL) as the overarching framework to address the scarcity of PII-labeled data in social media. Finally, we summarize previous studies to learn how syntactic patterns can help extract PII without pre-trained word embeddings.

### A. PII Exposure on Social Media

Internet users often expose various types of PII attributes, such as their names, birthdates, and geolocations, when using social media [1]. Such exposed PII, embedded in users' posts, profiles, behaviors, or social network relationships, can be easily exploited by adversaries and cyber criminals and lead to severe losses. In particular, social media posts published on social media platforms are a major and the most common source of PII exposure [2]. Thus, it is crucial to analyze social media posts to help raise users' privacy risk awareness.

Most existing research analyzes PII exposure in social media posts manually. For example, many studies leverage metrics or heuristic rules defined by experts to measure the extent of PII exposure in a user's posts [7], [8]. Although some studies leverage automated techniques (e.g., Naïve Bayes, LDA), they still require human efforts for feature engineering, which can be inefficient for the enormous social media data [9], [10]. This issue motivates us to review the information extraction (IE) literature as a viable approach to extract PII on social media automatically and efficiently.

### B. Deep Learning for Information Extraction (IE)

Information Extraction (IE) is a process of extracting target information (e.g., location, name, birthdate) from unstructured textual data automatically [11]. In recent years, deep learning has led to significant achievements in IE by alleviating the need for feature engineering. DL-based IE models usually have three major components: input representations, context encoder, and label encoder, as described next.

As with most Natural Language Processing (NLP) tasks, deep learning-based IE usually requires transforming texts into machine-readable input representations, often known as word embeddings [12]. Recent studies indicate that pre-trained word embeddings, such as Word2Vec, GloVe, and fastText, are critical for deep learning models, and often lead to better performance in the IE task [13]. To obtain more fine-grained inputs, previous studies have considered enhancing word representations with character-based representations, which can promote sensitivity to the spelling of words and enable capturing words' morphological features to help IE [14]. Recurrent Neural Networks (RNNs) and Convolutional Neural Networks (CNNs) are two widely-used architectures for extracting character-level representations in previous IE research [15]-[18].

Context encoder is a critical component in IE models to capture contextual dependencies of the inputs. RNNs and Transformers are two commonly used context encoder architectures [15]-[19]. Unlike RNNs that only focus on the local context (i.e., surrounding words) of a sequence, Transformers also consider global context (i.e., entire sentence). Thus, transformer-based models often outperform RNN-based models (e.g., Bi-directional Long Short-Term Memory (Bi-LSTM)) in IE [19].

Label encoder is usually the last component of IE models, and it aims to predict classes of words (e.g., location, birthdate) in a given sequence. Conditional Random Field (CRF) is the most common choice of label encoder in prior research [15]-[18]. CRF considers sequential relationships of the outputs, which can improve IE performance since dependencies across outputs often exist in IE (e.g., IOB (Inside, Outside, Beginning) tagging) [20].

Overall, we make two observations from the literature. First, most extant DL-based IE models require large-scale labeled data for training. However, PII-labeled social media data is difficult to obtain due to privacy concerns [1]. Second, existing methods rely heavily on pre-trained word embeddings. This can limit PIIE because traditional pre-trained word embeddings often lack PII information. Specifically, most PII cannot be captured by fixed pattern matching due to variation in forms. Syntactic patterns can capture words' role (e.g., subject, object, modifier) in a sentence to guide PIIE without relying on embeddings [15]. As a result, to address the reliance of existing models on large-scale labeled data, we review DTL as a viable solution. Besides, we review how syntactic patterns can help extract PII without pre-trained word embeddings.

### C. Deep Transfer Learning (DTL)

Deep Transfer Learning (DTL) aims to leverage a deep neural network architecture to extract and transfer latent source-domain knowledge to a target domain, thereby improving the performance of a task in the target domain. DTL can be classified into four categories: instances-based, mapping-based, adversarial-based, and network-based DTL [21]. Instance-based DTL utilizes selected instances in the source domain as supplements to the target domain. Mapping-based DTL maps instances from the source and target domains into a new data space as the training set. Adversarial-based DTL utilizes adversarial learning to find transferable features suitable for both domains. These three DTL paradigms usually require labeled data in the target domain [21]. In contrast, network-based DTL pre-trains a model in the source domain and transfers pre-trained layers to the target domain. This approach is suitable for PIIE since it does not require labeled data in the target domain (i.e., social media in this study), while PII-labeled public data can be used as the source domain. Therefore, DTL is expected to provide significant improvements for PIIE in the target domain.

### D. Syntactic Patterns for PIIE

Previous studies indicate that syntactic patterns are transferrable across domains, where the syntactic commonalities between domains help improve the performance of the target domain task [22]. Figure 1 illustrates the state-of-the-art IE model, Dependency-Guided LSTM-CRF, which is one of the few studies that consider syntactic patterns [15]. It concatenates word embeddings and the outputs of character Bi-LSTM as the input representations, which are fed into a Bi-LSTM to gain contextual information. A CRF layer is then used to predict the label of each word in a sentence. This model is suitable for PIIE, because syntactic patterns help capture words' roles in a sentence to guide PIIE with no embeddings.

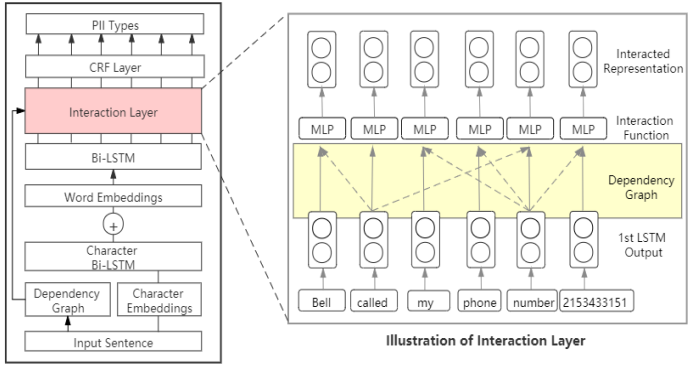

Fig. 1. Dependency-Guided LSTM-CRF model for IE [15]

This model leverages an *interaction layer* to capture syntactic patterns. At each position in the interaction layer, the outputs of the first Bi-LSTM propagate along with its directly related words (i.e., parent words) in a dependency graph to the next layer, and a Multilayer Perceptron (MLP) captures the interactions between syntactically related words. This method maps the dependency graph to sequential dependency relationships as a simplified approach to utilize syntactic patterns for IE.

Despite the novelty of the interaction layer, the model only focuses on sequential dependencies when learning syntactic patterns. To better capture the dependencies, syntactic patterns of a sentence can be represented in a graph structure (e.g., dependency graphs), where the nodes are words and edges are syntactic relationships between words [23]. Thus, dependency-guided LSTM-CRF overlooks the graph structure of syntactic patterns in a sentence and may cause information loss when extracting information.

To this end, we identify GCN as a possible solution [24]. The core idea of GCN is to generate nodal representations by propagating the information from the neighboring nodes, where the neighbors can be syntactically related words. In particular, GCN can integrate the syntactic patterns and word representations by taking them as structural features and nodal features, respectively. Syntactic patterns capture grammatical roles of words, and nodal representations reveal the meaning of words (i.e., semantics). Therefore, GCN can enhance the PIIE performance with more information.

### III. RESEARCH GAPS AND QUESTIONS

We identify two research gaps from the literature review. First, state-of-the-art DL-based IE models require large-scale PII-labeled data for training, while such data is rare in the social media domain. Second, syntactic patterns are critical to PIIE since they help extract PII without pre-trained word embeddings. However, existing methods primarily focus on sequential dependencies of syntactic patterns and overlook the graph structure of syntactic patterns. These gaps motivate the following research questions:

- How can we leverage DTL to conduct PIIE in social media to address the lack of PII-labeled data?
- How can we utilize GCN to capture syntactic patterns to alleviate the need for pre-trained word embeddings and improve PIIE performance?

### IV. RESEARCH DESIGN

We propose a social media PIIE framework, as illustrated in Figure 2. It comprises of three primary components: Data Collection, Data Pre-processing, and Exposed PII Extraction.

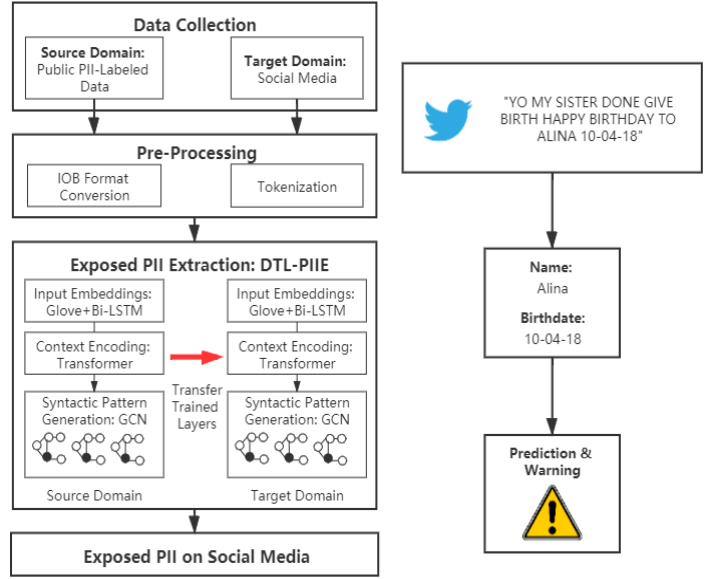

Fig. 2. Proposed social media DTL-PIIE framework

In Data Collection, we leverage public datasets that are manually labeled with PII as the source domain and collect posts from social media as the target domain. The collected data is converted into the IOB format and tokenized so that it can be used to train and test our proposed model in the IE task. In Exposed PII Extraction, we propose a novel deep learning-based IE model, DTL-PIIE, consisting of three main components. First, the input embeddings are generated at the word and character levels using GloVe and character Bi-LSTM, respectively. Then, we utilize Transformers as the context encoder. Finally, we leverage GCN to incorporate input representations and syntactic patterns into a single embeddings for PIIE. Next, we explain our proposed DTL-PIIE model in more detail.

As shown in Figure 3, our proposed DTL-PIIE model consists of three steps. First, the model is trained on the source domain. Second, we transfer Character Bi-LSTM, Transformer, and PII-GCN layers (shaded layers on the left side of Figure 3) to the target domain (shaded layers on the right side of Figure 3). Third, the transferred layers are reused, and the CRF layer is trained in the target domain. The transferred layers can also be fine-tuned if sufficient training data is available.

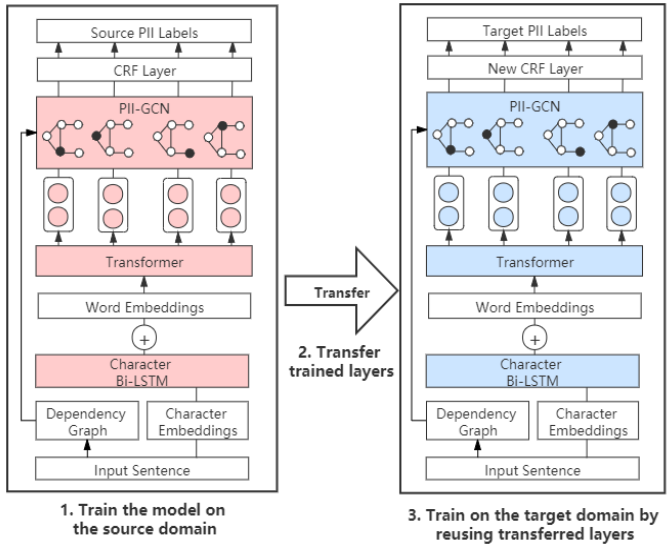

Fig. 3. Proposed DTL-PIIE model. **Note:** Red represents layers before transferred and blue layers are transferred, and other layers are fixed.

Specifically, we transfer Character Bi-LSTM, because it can capture PII's morphological features (e.g., suffix '-son' often indicates an English last name). Besides, Transformer is transferred since it can capture the contexts of PII. For instance, "*Dr.*" and "*live in*" as the left context are often followed by a name and a location, respectively. Lastly, we transfer PII-GCN to enhance the PIIE performance with syntactic patterns. PII-GCN serves to learn syntactic patterns of sentences that contain PII, which is the main novelty of our model.

Figure 4 illustrates the key PII-GCN component in our proposed DTL-PIIE model. The syntactic relationships between words are indicated by the dependency graph, where the nodes are words in the sentence, and the edges are the dependency relation between $w_i$ and $w_j$, if exits.

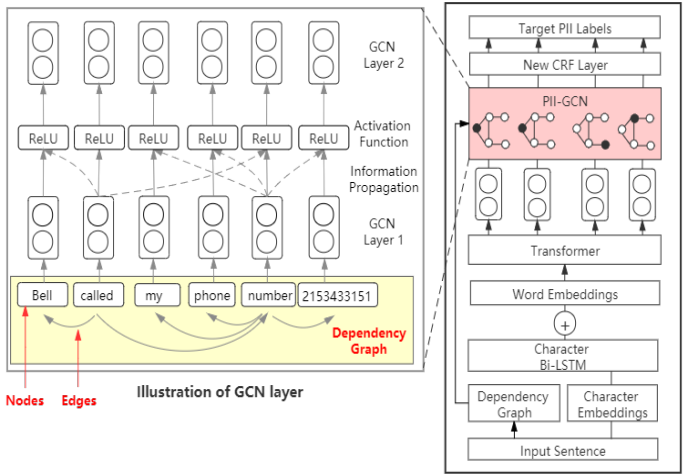

Fig. 4. Illustration of PII-GCN in the Proposed DTL-PIIE model. **Note:** Red highlights our technical novelty; yellow box is the dependency graph.

The graph guides PII-GCN to propagate the word representations based on syntactically related words. The information from the neighboring words are incorporated to form the new representation of a word. Specifically, a GCN model at layer $l$ is denoted as [24]:

$$H_l = f(H_{l-1}, A) = \sigma(\tilde{A} H_{l-1} W_{l-1}) \quad (1)$$

where $W_{l-1}$ is a trainable weight matrix and $\sigma$ is ReLu activation. At the first layer, $H_0 = X$, which is the matrix of nodal features. $\tilde{A}$ is the normalized adjacency matrix to avoid vanishing or exploding gradients:

$$\tilde{A} = \bar{D}^{-\frac{1}{2}} \bar{A} \bar{D}^{-\frac{1}{2}} \quad (2)$$

$$\bar{A} = I + A \quad (3)$$

where $I$ is the identity matrix, and $\bar{D}$ is the degree matrix of $\bar{A}$. $\bar{A}$ represents adding an identity matrix to the adjacency matrix $A$ to allow a self-loop to each node. PII-GCN generates nodal representations of each word by propagating the information from the neighboring nodes, where the neighbors are syntactically related words in the dependency graph. This propagation process enables the model to integrate word representations (i.e., nodal features) and dependency graphs (i.e., structural features) into a single embedding to incorporate syntactic patterns.

## V. EXPERIMENTS

### A. Research Testbed Development

We developed a comprehensive research testbed by leveraging multiple public datasets and collecting raw data from social media to enable our experiments. For the source domain, we combined six public datasets that have been labeled and used in prior PII research [4], [5], [25]-[27]. These datasets were chosen because they are publicly available and developed by researchers.

For the target domain, as discussed, labeled data is difficult to obtain. Hence, we collected a social media dataset comprising 8,000 tweets and formed a panel of two privacy experts to manually annotate these tweets. Specifically, the annotators labeled a word if it belongs to any of the following 7 PII categories: Age, Contact, Date, ID, Location, Name, and Profession. The labels were confirmed only if both annotators agreed, and the agreement rate was above 90%.

The labeled data was then pre-processed. Specifically, to normalize the corpus, we converted all characters to lowercase and leveraged WordNet Lemmatizer to unify the words with different spellings. Furthermore, we removed non-alphanumeric characters since PII usually does not contain symbols, while dash (-), slash (/), and at symbol (@) were preserved, since they could be found in birthdates and email addresses, respectively. Lastly, both the source-domain and target-domain data was tokenized and converted into the IOB format for the ease of training and testing the models. We summarize our research testbed in Table I.

TABLE I. SUMMARY OF RESEARCH TESTBED

| PII Category | Source Domain | | | | | | Total | Target Domain |
| --- | --- | --- | --- | --- | --- | --- | --- | --- |
| | i2b2 2014 | CoNLL 2003 | GMB-1.0.0 | WNUT 17 | Broad Twitter Corpus | Resume Entities for NER | | Self-collected from Twitter |
| Age | 1,997 | - | - | - | - | 220 | 2,217 | 786 |
| Contact | 541 | - | - | - | - | 220 | 761 | 912 |
| Date | 12,487 | - | 12,786 | - | - | 220 | 25,493 | 1,026 |
| ID | 1,506 | - | - | - | - | - | 1,506 | 452 |
| Location | 4,580 | 10,645 | 58,388 | 213 | 3,114 | 220 | 77,160 | 362 |
| Name | 7,348 | 10,059 | 20,366 | 894 | 5,271 | 220 | 44,158 | 3,062 |
| Profession | 413 | - | - | - | - | 220 | 633 | 188 |
| # of document | 1,304 | 1,393 | 9,999 | 2,295 | 9,551 | 220 | 24,762 | 8,000 |

## B. Experiment Results

We evaluated our proposed DTL-PIIE model with two experiments. First, we compared our model with state-of-the-art DL-based IE models, including Dependency-Guided LSTM-CRF. Second, we conducted an ablation analysis to examine the contribution of different components in our model. To assess the models' performance, well-established metrics, including Accuracy, Precision, Recall, and F1 Score were used. Hyper-parameters were tuned through 5-fold cross-validation. Experiments were conducted on a single Microsoft Windows 10 Pro Server with 128GB of RAM, an Nvidia GeForce GTX 2080 GPU, and an Intel CPU at 2.60 Gigahertz (GHz). Table II summarizes our experiment results.

TABLE II. STOLEN CREDIT/DEBIT CARD COLLECTION RESULTS

| Domain | Method | A | P | R | F1 |
| --- | --- | --- | --- | --- | --- |
| Source | Bi-LSTM + CRF (no transfer) | 95.3% | 92.6% | 89.0% | 91.6% |
| | Transformer + SoftMax | 94.1% | 93.0% | 90.2% | 92.5% |
| | Dependency-Guided LSTM-CRF | 97.8% | 94.5% | 93.3% | 94.0% |
| | **Source Domain DTL-PIIE** | **98.1%** | **95.9%** | **94.1%** | **94.6%** |
| Target (Twitter) | Dependency-Guided LSTM-CRF | 59.3% | 61.0% | 54.6% | 57.2% |
| | Transformer + SoftMax | 64.9% | 67.2% | 60.4% | 65.1% |
| | Bi-LSTM + CRF + Transfer | 68.7% | 68.3% | 62.5% | 67.4% |
| | **DTL-PIIE (Ours)** | **73.2%** | **72.5%** | **65.2%** | **71.1%** |

The state-of-the-art DL-based IE models that we used as the benchmark are Bi-LSTM + CRF + Transfer [16], Transformer + SoftMax [19], and Dependency-Guided LSTM-CRF [15]. As seen in Table II, in the source domain, our model achieved the best performance in all metrics, surpassing the Dependency-Guided LSTM-CRF with 0.6% higher F1 score (94.6% and 94.0%). In the target domain, the superiority of our model over the benchmark models was more salient. Compared with Bi-LSTM + CRF + Transfer, our model achieved 13.9% increase in F1 score (71.1% and 67.4%).

The superior performance of our model can be mainly attributed to three main components: a Transformer-based context encoder, a PII-GCN for learning syntactic patterns, and DTL for knowledge transfer. Transformer leveraged global context to emphasize words that helped identify PII in social media posts. For example, a user can mention "phone number" at the beginning while leaving the number at the end of a post. In this case, "phone number" is not in the local contexts of the PII (i.e., the number) due to the long distance between them, but Transformer can capture the global contexts and extract the PII.

In comparison with Dependency-guided LSTM-CRF that only used sequential dependencies of syntactic patterns, our model leveraged PII-GCN to incorporate the graph structure of syntactic patterns. The sequential dependencies of syntactic patterns only consider directly related words in the syntax, whereas in social media, words directly related to PII can often be written in irregular forms (e.g., bd stands for birthdate). These words also cannot be represented by pre-trained word embeddings, and their syntax-related words are further needed to guide PIIE. Hence, our model with PII-GCN considers all words related to PII in the dependency graph and outperforms prior models in PIIE. Last but not least, our results suggest that DTL can effectively generalize the global contexts and syntactic patterns learned from publicly available PII data to the social media domain to improve PIIE performance.

## C. Experiment Results - Ablation Analysis

Next, we conducted ablation analysis to examine the contribution of different components in our model. Specifically, in the source domain, we dropped GCN and Dependency Graph as the first setting, since they are the most critical components of our model, and dropping them can test the contribution of incorporating syntactic patterns to PIIE. Second, we replace Transformer with Bi-LSTM that has been used by prior research to test whether Transformer performs better than Bi-LSTM as the context encoder. In the target domain, we drop the whole DTL component as an additional setting to gain a better understanding of DTL's contribution in our model. The results of the ablation analysis are summarized in Table III.

As shown in Table III, PII-GCN and Dependency Graph increased F1 score by 3.6% (from 91.0% to 94.6%) in the source domain and by 3.9% (from 91.0% to 94.6%) in the target domain. Moreover, using Transformer as the context encoder attained better performance than using Bi-LSTM with 0.8% and 2.1% higher F1 scores in the source domain and target domain, respectively. This suggests that unlike Transformer that only uses word embeddings to capture word representations, PII-GCN + Dependency Graph leverage syntactic patterns to focus on neighboring words when learning the representations of PII. These representations turn out to be more effective in PIIE where pre-trained word embeddings are not often available. Furthermore, in the target domain, DTL improved the F1 score by 11.4% (from 59.7% to 71.1%), indicating that DTL can effectively improve PIIE performance when PII-labeled data is scarce in social media.

TABLE III. STOLEN ACCOUNT CREDENTIAL COLLECTION RESULTS

| Domain | Dropped Layer (s) | A | P | R | F1 |
|---|---|---|---|---|---|
| Source | PII-GCN + Dependency Graph | 92.4% | 92.5% | 89.7% | 91.0% |
| | Transformer | 97.6% | 95.8% | 93.9% | 93.8% |
| | **None (Source Domain DTL-PIIE)** | **98.1%** | **95.9%** | **94.1%** | **94.6%** |
| Target (Twitter) | DTL | 62.0% | 62.5% | 57.8% | 59.7% |
| | PII-GCN + Dependency Graph | 69.4% | 69.0% | 62.6% | 67.2% |
| | Transformer | 72.1% | 71.3% | 62.5% | 69.0% |
| | **None (DTL-PIIE)** | **73.2%** | **72.5%** | **65.2%** | **71.1%** |

## VI. CONCLUSION AND FUTURE DIRECTIONS

In this study, we aim to automatically extract PII in social media to raise users' privacy awareness. While exposed PII can be misused by cyber criminals and cause severe losses, PIIE can help users take protective measures. Nonetheless, automated PIIE is challenging because most prevailing DL-based IE models rely heavily on large-scale labeled data and pre-trained word embeddings, both of which are rare for social media PII. We proposed DTL-PIIE to utilize transfer learning to address the data scarcity problem as well as GCN to capture syntactic patterns for better PIIE in social media data. Our proposed framework can facilitate various applications, such as the prediction of PII misuse and privacy risk assessment, to improve users' privacy risk awareness and encourage them to take protective measures. Future research can focus on exposed PII in non-textual contents (e.g., photos, videos) to provide a wider range of protection for internet users. Also, researchers may leverage DTL-PIIE to study how users react when they become aware of their PII exposure, and how such reactions vary in different at-risk populations.

ACKNOWLEDGMENTS

This material is based upon work supported by the National Science Foundation (NSF) under Secure and Trustworthy Cyberspace (SaTC) (grant No. 1936370).

REFERENCES

[1] J. Isaak and M. J. Hanna, "User data privacy: Facebook, Cambridge Analytica, and privacy protection," *Computer (Long. Beach. Calif).*, vol. 51, no. 8, pp. 56–59, 2018.

[2] B. Krishnamurthy and C. E. Wills, "On the leakage of personally identifiable information via online social networks," in *Proceedings of the 2nd ACM workshop on Online social networks*, 2009, pp. 7–12.

[3] C. Cadwalladr and E. Graham-Harrison, "Revealed: 50 million Facebook profiles harvested for Cambridge Analytica in major data breach," *Guard.*, vol. 17, p. 22, 2018.

[4] A. Stubbs, C. Kotfila, and Ö. Uzuner, "Automated systems for the de-identification of longitudinal clinical narratives: Overview of 2014 i2b2/UTHealth shared task Track 1," *J. Biomed. Inform.*, vol. 58, pp. S11–S19, 2015.

[5] J. Bos, V. Basile, K. Evang, N. J. Venhuizen, and J. Bjerva, "The groningen meaning bank," in *Handbook of linguistic annotation*, Springer, 2017, pp. 463–496.

[6] F. Lin et al., "Linking personally identifiable information from the dark web to the surface web: A deep entity resolution approach," in *2020 International Conference on Data Mining Workshops (ICDMW)*, 2020, pp. 488–495.

[7] R. G. Pensa, G. Di Blasi, and L. Bioglio, "Network-aware privacy risk estimation in online social networks," *Soc. Netw. Anal. Min.*, vol. 9, no. 1, p. 15, 2019.

[8] L. Yu et al., "My friend leaks my privacy: Modeling and analyzing privacy in social networks," in *Proceedings of the 23nd ACM on Symposium on Access Control Models and Technologies*, 2018, pp. 93–104.

[9] E. Aghasian, S. Garg, and J. Montgomery, "An automated model to score the privacy of unstructured information—Social media case," *Comput. Secur.*, vol. 92, p. 101778, 2020.

[10] D. Keküllüoglu, W. Magdy, and K. Vaniea, "Analysing privacy leakage of life events on twitter," in *12th ACM Conference on Web Science*, 2020, pp. 287–294.

[11] K. Adnan and R. Akbar, "An analytical study of information extraction from unstructured and multidimensional big data," *J. Big Data*, vol. 6, no. 1, pp. 1–38, 2019.

[12] T. Schnabel, I. Labutov, D. Mimno, and T. Joachims, "Evaluation methods for unsupervised word embeddings," in *Proceedings of the 2015 conference on empirical methods in natural language processing*, 2015, pp. 298–307.

[13] M. Habibi, L. Weber, M. Neves, D. L. Wiegandt, and U. Leser, "Deep learning with word embeddings improves biomedical named entity recognition," *Bioinformatics*, vol. 33, no. 14, pp. i37–i48, 2017.

[14] G. Lample, M. Ballesteros, S. Subramanian, K. Kawakami, and C. Dyer, "Neural architectures for named entity recognition," *arXiv Prepr. arXiv1603.01360*, 2016.

[15] Z. Jie and W. Lu, "Dependency-guided LSTM-CRF for named entity recognition," in *Proceedings of the 2019 Conference on Empirical Methods in Natural Language Processing and the 9th International Joint Conference on Natural Language Processing (EMNLP-IJCNLP)*, 2019, pp. 3862-3872.

[16] J. Y. Lee, F. Dernoncourt, and P. Szolovits, "Transfer learning for named-entity recognition with neural networks," in *Proceedings of the Eleventh International Conference on Language Resources and Evaluation (LREC 2018)*, 2018.

[17] S. Moon, L. Neves, and V. Carvalho, "Multimodal named entity recognition for short social media posts," *arXiv Prepr. arXiv1802.07862*, 2018.

[18] Y. Nie, Y. Tian, Y. Song, X. Ao, and X. Wan, "Improving named entity recognition with attentive ensemble of syntactic information," *arXiv Prepr. arXiv2010.15466*, 2020.

[19] J. Devlin, M.-W. Chang, K. Lee, and K. Toutanova, "Bert: Pre-training of deep bidirectional transformers for language understanding," *arXiv Prepr. arXiv1810.04805*, 2018.

[20] J. Lafferty, A. McCallum, and F. C. N. Pereira, "Conditional random fields: Probabilistic models for segmenting and labeling sequence data," 2001.

[21] C. Tan, F. Sun, T. Kong, W. Zhang, C. Yang, and C. Liu, "A survey on deep transfer learning," in *International conference on artificial neural networks*, 2018, pp. 270–279.

[22] B. A. Galitsky, "Transfer learning of syntactic structures for building taxonomies for search engines," *Eng. Appl. Artif. Intell.*, vol. 26, no. 10, pp. 2504–2515, 2013.

[23] D. Cai and W. Lam, "Graph transformer for graph-to-sequence learning," in *Proceedings of the AAAI Conference on Artificial Intelligence*, 2020, vol. 34, no. 05, pp. 7464–7471.

[24] T. N. Kipf and M. Welling, "Semi-supervised classification with graph convolutional networks," *arXiv Prepr. arXiv1609.02907*, 2016.

[25] E. F. Sang and F. De Meulder, "Introduction to the CoNLL-2003 shared task: Language-independent named entity recognition," *arXiv Prepr. cs/0306050*, 2003.

[26] L. Derczynski, K. Bontcheva, and I. Roberts, "Broad twitter corpus: A diverse named entity recognition resource," in *Proceedings of COLING 2016, the 26th International Conference on Computational Linguistics: Technical Papers*, 2016, pp. 1169–1179.

[27] L. Derczynski, E. Nichols, M. van Erp, and N. Limsopatham, "Results of the WNUT2017 shared task on novel and emerging entity recognition," in *Proceedings of the 3rd Workshop on Noisy User-generated Text*, 2017, pp. 140–147.